\documentclass{article}
\newcommand{\mfrac}[2]
{\mbox{\footnotesize$\displaystyle\frac{\raisebox{-0.1em}
{\mbox{$#1$}}}{#2}$}}
\newcommand{\pot}{{\footnotesize $\mbox{\footnotesize $[$}\, U\, 
\mbox{\footnotesize $]$}$}}

\begin{document}
\centerline{\large\scshape On the Dubrovin equations for the
  finite-gap potentials}

\vskip 0.3cm \centerline{{\it Y.\,V. Brezhnev}\footnote {{\em
      Kaliningrad State University\/}; \ e-mail:
    \texttt{brezhnev@mail.ru}}}

\noindent
Until recently, the following question has not been considered: how to
generalise constructions, related to the Dubrovin's equations (DE) in
finite-gap potential theory for the Schr\"odinger operator, into
arbitrary spectral problems.  We have in mind the equations of the zeros
of the $\Psi$-function and the trace formulas.  In spite of physical
interpretation of these objects as analogs of the scattering data,
the generalizations are not clear, or may even be absent. This
problem has an independent interest. For example, the well known
Novikov's equations, appearing in a general theory of finite-gap
integration, are (to all appearences) completely integrable
finite-dimensional dynamical systems.  A reduction of the DE to the
Jacobi inverse problem demonstates the Liouville's integrability of these
equations. Note that if an algebraic curve is a cover over an elliptic
curve, then one can use the trace formulas to produce solutions in
elliptic functions, new finite-gap potentials and some applications to
nonlinear integrable partial differential equations.

In a recent note [2] an universal feature of the finite-gap potentials
was revealed: they form a class, which admits an integration of the
spectral problem by quadratures. It was shown there how to obtain all
the ingredients of the straight spectral problem: the $\Psi$-formula,
algebraic curve, Novikov's equations and their integrals. On the other
hand, as soon as $\Psi$ is known, it is natural to expect that the
equations for its zeroes $\gamma_k^{}(x)$ may be written using simple
arguments. This can be done, and we show algorithmically how to solve
the problem with the appearance of related objects: trace formulas and
the Abel transformation.  We do not discuss here a separate question
about an exact (or one-to-one) 
correspondence between the following two constructions:
\begin{itemize} 

\item an algebraic curve and the
divisor of the zeroes $\{\gamma_k^{}\}$; 

\item boundary
conditions of the Dirichlet type $\Psi(x_{\mbox{\scriptsize
    o}})=\Psi(x_{\mbox{\scriptsize o}} + \mbox{\small $\Omega$})=0$;

\end{itemize} 
if to consider these topics 
in the context of recovering the
potential from scattering data on an interval for arbitrary spectral
problems.  The set $\{\gamma_k^{}\}$ is considered as a starting
point. We demonstrate our approach on an example of a scalar operator
$\lambda$-pencil of the third order.

\begin{equation}
  \Psi''' + u(\lambda;x)\, \Psi'+v(\lambda; x)\, \Psi=0,
\end{equation}
where $u,\,v$ are rational functions of $\lambda$ with poles
independent of $x$. We call their potential \pot\ and fix it to be a
finite-gap potential. Thus, $\Psi$ can become the multi-point
Baker--Akhiezer function [3].  An operator pencil, commuting with
equation (1) (that is, the second equation on the common
eigenfunction) in general has the form:
$$
A(\lambda;x)\,\Psi''+ B(\lambda; x)\,\Psi'+ C(\lambda;x)\,\Psi =
\mu\, \Psi, \qquad C=\mfrac23\, u^{\mathstrut}\,A -\mfrac13\,A''- B',
$$
where $A$, $B$ are polynomials in $\lambda$ and differential
polynomials in \pot. Following [2], one obtains an algebraic curve
$W(\mu, \lambda)=0$ and two necessary representations for the
$\Psi$-function.
\begin{equation}
  \mfrac{\Psi'}{\Psi}= \mfrac{(A'+B)\,(\mu-C)+A\,C'-v\,A^2}{A\,\mu-F}=
  -3\,\mfrac{(\mu^2+Q)\,A^2+A\,F\,\mu+F^2}{A\,\Pi}+\mfrac{A'+B}{A},
\end{equation}
$$
\begin{array}{c}
  -3\,F=A\,A''+u\,A^2_{{}_{\mathstrut}}+3\,B\,(A'+B),\\
  \Pi(\lambda;x)=(3\,v-u')\,A^{3^{\mathstrut}}+(2\,A'''+
  3\,B'' + 3\,u\,B+2\,u\,A')\,A^{2^{\mathstrut}}+\\
  +3\,A\,B\, (A''+B' )+3\,B^2\,(A'+B).
\end{array}
$$
The second expression for the $\Psi$-function in (2) was obtained
by conversion of the first expression in (2) into a polynomial as a
rational function in $\mu$ with the help of the equation of the curve:
$W(\mu, \lambda) \equiv\mu^3+Q(\lambda)\,\mu+R(\lambda)=0$.  A sum of
zeroes $\{\gamma_k^{}\}$ on all sheets of the Riemann surface
$W(\mu,\lambda)=0$ is defined by the poles of the expression
$\Psi'\!/\Psi$ and factorises the denominator
$\Pi=a\!\cdot\!(\lambda-\gamma_1^{}) \cdots (\lambda-\gamma_n^{})$.
In turn, the denominator of the first formula in (2) defines the
second coordinate (number of a sheet) on the curve of the
$\Psi$-function zero: $\mu_k^{}=\mu(\gamma_k^{}(x))$.  Thus,
$F(\lambda;x)$ leads to $\mu_k^{}(x)\,A(\gamma_k^{};x)$ when $\lambda
\to \gamma_k^{}$.  Using this fact and taking the passage to the limit
$\lambda \to \gamma_k^{}$ in the second formula (2) one obtains
analogs of the DE.
\begin{equation}
  \displaystyle\gamma'_k =3\,A(\gamma_k^{};x)\,\frac{3\,\mu_k^2+
    Q(\gamma_k^{})} {\displaystyle a \prod _{j \ne
      k}(\gamma_k^{}-\gamma_j^{})}, \qquad
  \mu'_k=-3\,A(\gamma_k^{};x)\, \frac{Q'(\gamma_k^{})\,\mu_k^{}+
    R'(\gamma_k^{})}
  {\displaystyle a \prod _{j \ne k}(\gamma_k^{}-\gamma_j^{})},
\end{equation}
$$
\mu_k^{}=\frac{F(\gamma_k^{};\,x)}{A(\gamma_k^{};\,x)}.
$$
Up to this point we have not placed any restrictions, therefore the
equations (3) hold for any operator pencil (1) and the method of
derivation is spread to the higher orders without any changes.  The
first system of equations in (3) recently appeared in [4] 
(not taking into account a misprint in eq. (5.34) on p.\,852)
as an
example of the Boussinesq equation.  However proofs and  examples
were not presented.  
  We would like to emphasise two important circumstances not touched on in [4]. 
\begin{enumerate}
\item
The DE must be reduced to an autonomous form, but in the form (3),
they contains a potential in $A$-function;

\item The potential \pot\ is
expressed in terms of $\Psi$-function (i.e. $\Theta$).  
\end{enumerate}
Based on this, we formulate a problem: {\em which operator
  $\lambda$-pencils allow us to recover uniquely a finite-gap
  potential by zeros-coordinates of the $\Psi$-function\/}?  {\em If
  yes, whether it will be an Abelian function on a Jacobian of a
  curve\/}?  Note, that the Schr\"odinger operator (excepting the
simple modifications of a $2 \times 2$-Dirac-operator) is the only
known example, where this question disappears due to the first trace
formula containing $\sum \gamma_k^{}$.  We call it a {\em central\/} trace
formula as all others are its consequence. 
But an arbitrary Abelian function is a symmetrical
combination of the upper bounds of the Abelian integrals in the Jacobi
inverse problem.  Thus, coordinates $(\gamma_k^{},\,\mu_k^{})$ must be
considered as having a same importance, and possibly both will appear
in the trace equalities.  This requires the appearance of the second
group of the equations (3). Note, that the third formula in (3) is the
consequence of an algebraic curve $W(\mu,\, \lambda)=0$ and the
$\Psi$-function formula (2), i.e. the second coordinate $\mu_k^{}$ is
always determined.  It is not difficult to find examples, when
solution of the point 1) is not clear (or even impossible) and
a general recipe for the solution of the  problem is not known
at present.

Without specification of the $A,\,B$-polynomials and the potential
\pot, traces and the Jacobi inverse problem can not be written,
because they entirely define the structure of a curve and essential
singularities of the $\Psi$-function.  Therefore, let us consider an
example (genus $g=4$).
$$
\Psi'''+u(x)\,\Psi'=\lambda\,\Psi,\qquad
(3\,u'-9\,\lambda)\,\Psi''+(u^2-u''+\alpha)\,\Psi'-
6\,\lambda\,u\,\Psi=\mu\,\Psi,
$$
\begin{equation} 
  W(\mu,\,\lambda): \qquad \mu^3 +(27\,\alpha\,\lambda^2+E_2^{})\,\mu
  +729\,\lambda^5+81\,E_1^{}\,\lambda^3+ E_3^{}\,\lambda=0.
\end{equation}
A factorisation of the $\Pi$-polynomial yields only the formula
$2\,u'= 3\sum \gamma_k^{}$. But combining it with the expression of
the integrals $E_{1,3}^{}(u,u',\ldots, u^{\mbox{{\sc\tiny (iv)}}})$ and
formula (3) we obtain, as an answer to the above question, three
versions of the central trace formula:
$$
u =  -\mfrac{3}{2\,\alpha} \,\Big\{E_1^{} +\sum_{k=1}^{4}\,
(\gamma_k'''+\mfrac{15}{2}\,
\gamma_k^2)\Big\},  \qquad 
u = \mfrac16\,\sum_{k=1}^4
\,\gamma_k^{}\,\mu_k^{}\,\frac{\sum\limits^4\gamma_j^{} -2\,\gamma_k^{}}
{\displaystyle\prod _{j \ne k}(\gamma_k^{}-\gamma_j^{})},
$$
$$
u=\frac{E_3^{}+3^6\, \gamma_1^{}\,\gamma_2^{}\,\gamma_3^{}\,\gamma_4^{}}
{6\,E_1^{}}.
$$
The integral form of  equations (3) as the Jacobi inverse problem
and the base of the holomorphic differentials for the trigonal 
curve (4) have the form
\begin{equation}
\left\{
\begin{array}{ll}
  \displaystyle\sum_{k=1}^4
  \!\!\int\limits_{}^{(\gamma_k^{},\,\mu_k^{})}
  \!\!\!\!\frac{d\lambda}
  {3\,\mu^2+Q(\lambda)}=a_1^{},&\displaystyle\quad \sum_{k=1}^4
  \!\!\int\limits_{}^{(\gamma_k^{},\,\mu_k^{})} \!\!\!\!
  \frac{\lambda\,d\lambda} {3\,\mu^2+Q(\lambda)}=a_2^{},
  \\
  \displaystyle \sum_{k=1}^4
  \!\!\int\limits^{(\gamma_k^{},\,\mu_k^{})} \!\!\!\!
  \frac{\mu\,d\lambda} {3\,\mu^2+Q(\lambda)}=a_3^{}, &
  \displaystyle\quad \sum_{k=1}^4
  \!\!\int\limits^{(\gamma_k^{},\,\mu_k^{})}
  \!\!\!\!\frac{\lambda^2\,d\lambda}
  {3\,\mu^2+Q(\lambda)}=a_4^{}-\mfrac{1}{81}\,x,
\end{array}
\right.
\end{equation}
where $Q(\lambda)=27\,\alpha\,\lambda^2+E_2^{}$.  
The following nontrivial example is taken from the hierarchy of 
Boussinesq's equation without reduction.
$$
\Psi'''+u\,\Psi'+v\,\Psi=\lambda\,\Psi, \quad 
u\,\Psi'''+(3\,\lambda+v-u'+\alpha)\,\Psi'+\left(\mfrac23\,u''-v'+\mfrac23\,u^2
\right)
\Psi=\mu\,\Psi,
$$
$$
A=u,\qquad B=3\,\lambda+v-u'+\alpha,  \qquad \mbox{{\small genus}\ \ $g=3$}.
$$
The second coordinate $\mu_k^{}$ read as
$$
\mu_k^{}=-\frac{27\,\gamma_k^2+9\,(2\,v-u'+2\,\alpha)\,\gamma_k^{}
+u\,u''+u^3+3\,(\alpha+v)\,(\alpha+v-3\,u')}{3\,u}.
$$
The central trace formulas have the form:
$$
\frac9u = - \sum\limits_{k=1}^{3}\,\frac{\mu_k^{}}
{\displaystyle \prod\limits_{j \ne k}^{3}\,(\gamma_k^{}-\gamma_j^{})},
\qquad
v=\mfrac23\,u'-\alpha-\sum\limits_{k=1}^{3}\,\gamma_k^{}.
$$
We do not display here the correspondent curve,  autonomous DE
and their integral form type as (5). 
Note, the holomorphic Abelian differentials for an arbitrary algebraic curve
$W(\mu,\lambda)=0$ may be written in the canonical form
$d\,\omega=P(\mu,\lambda)\,W_{\!\mu}^{\mbox{-}\!1}\,d\lambda$ [5,
\S\,39], but we will not always arrive at the DE in such a way. It is
enough, to mention a counterexample $\Psi'' =\lambda\,u(x)\,\Psi$,
where neither the straight analogs of the DE nor trace formulas occur.
But, in this example, the $\Psi$-function is not a classical
Baker--Akhiezer function with the asymptotic behaviour
$\sim\exp(k\,x)$ [3], although a commuting operator, curve and
$\Psi$-formula are written easily.

The author thanks Dr. N.\,Ustinov for the discussions and acknowledges
the financial support of the Russian Basic Research Foundation
(00--01--00782).

\end{document}